\documentclass[aps,prl,twocolumn,groupedaddress,showpacs]{revtex4-1}
\usepackage{graphicx}
\usepackage{amsmath}
\usepackage{color}
\usepackage{float}
\usepackage{epstopdf}
\usepackage{soul,xcolor}
\usepackage[colorlinks=true,citecolor=blue,linkcolor=magenta]{hyperref}
\begin{document}

\title{Experimental Realization of non-Adiabatic Shortcut to non-Abelian Geometric Gates}

\author{Tongxing Yan$^{1,4}$}
\thanks{T.-X. Y. and B.-J. L. contributed equally to this work.}
\author{Bao-Jie Liu$^1$}
\thanks{T.-X. Y. and B.-J. L. contributed equally to this work.}
\author{Kai Xu$^2$, Chao Song$^2$ }
%\author{H. Wang$^{2,3}$}
%\email{hhwang@zju.edu.cn}
\author{Song Liu$^{1,3}$, Zhensheng Zhang$^{1,3}$}

\author{Hui Deng$^{6}$, Zhiguang Yan$^{6}$, Hao Rong$^{6}$}

\author{Keqiang Huang$^{5}$}

\author{Man-Hong Yung$^{1,3}$}
\email{yung@sustc.edu.cn }
\author{Yuanzhen Chen$^{1,3}$}
\email{chenyz@sustc.edu.cn}
\author{Dapeng Yu$^{1,3}$}
\affiliation{$^1$ Institute for Quantum Science and Engineering and Department of Physics, Southern University of Science and Technology, Shenzhen 518055, China,
$^2$ \mbox{Department of Physics, Zhejiang University, Hangzhou, Zhejiang 310027, China},
$^3$ \mbox{Shenzhen Key Laboratory of Quantum Science and Engineering, Shenzhen, 518055, China}
$^4$ \mbox{School of Physics, University of Chinese Academy of Sciences, Beijing, 100049, China},
$^5$ \mbox{Institute of Physics, Chinese Academy of Sciences, Beijing 100190, China},
$^6$ \mbox{CAS Center for Excellence and Synergetic Innovation Center in Quantum Information and Quantum Physics}, \mbox{University of Science and Technology of China, Hefei, Anhui 230026, China},
}

\date{\today }
\bibliographystyle{apsrev4-1}

%\pacs {75.30.Cr, 75.10.Jm, 75.40.Cx, 75.50.Ee}

\begin{abstract}

When a quantum system is driven adiabatically through a parametric cycle in a degenerate Hilbert space, the state would acquire a non-Abelian geometric phase, which is stable and forms the foundation for holonomic quantum computation (HQC). However, in the adiabatic limit, the environmental decoherence becomes a significant source of errors. Recently, various non-adiabatic HQC schemes have been proposed, but all at the price of increased sensitivity to control errors. Alternatively, there exist theoretical proposals for speeding up HQC by the technique of ``shortcut to adiabaticity" (STA), but no experimental demonstration has been reported so far, as these proprosals involve a complicated control of four energy levels simultaneously. Here we propose and experimentally demonstrate that HQC via shortcut to adiabaticity can be constructed with only three energy levels, using a superconducting qubit in a scalable architecture. With this scheme, all holonomic single-qubit operations can be realized non-adiabatically through a single cycle of state evolution. As a result, we are able to experimentally benchmark the stability of STA+HQC against NHQC in the same platform. The flexibility and simplicity of our scheme makes it also implementable on other systems, such as nitrogen-vacancy center, quantum dots, and nuclear magnetic resonance. 

\end{abstract}
%\pacs{03.67. Pp, 03.67.Lx, 85.25.Hv}

\vskip 0.5cm
\maketitle

\narrowtext

\emph{Introduction}.\textbf{--}In quantum information processing, logical operations are achieved by actively manipulating quantum evolutions, which may be dynamical and/or geometrical in nature. In the latter case, as the system Hamiltonian undergoes a cyclic evolution in a parameter space, a geometric phase accumulates~\cite{Berry1984,Wilczek1984PRL,Zarnadi1999PRL,Aharonov1987PRL,Anandan1988PRA}, which has been demonstrated to be intrinsically resilient to certain types of noises~\cite{Johansson2012PRA,Berger2013PRA,Yale2016Nature}. Therefore, it may be used to implement robust geometric quantum computation~\cite{Falci2000Nature,Jones2000Nat,Duan2001Science,Ditte2007PRA,Gerard2017PRL,Wang2001PRL,Zhu2003PRL,Sjoqvist2012NJP,Xu2012}. 

In particular, the non-commutativity nature of non-Abelian geometric phases~\cite{Wilczek1984PRL,Zarnadi1999PRL} makes it suitable for implementing quantum gates. Geometric quantum computation in this form is often referred as holonomic quantum computation (HQC). Originally, geometric quantum gates were constructed via adiabatic evolutions~\cite{Falci2000Nature,Duan2001Science,Ditte2007PRA,Gerard2017PRL}, which require a long runtime to avoid transitions among the instantaneous eigenstates of the Hamiltonian of interest. To overcome such a problem, non-adiabatic HQC schemes were proposed~\cite{Sjoqvist2012NJP,Xu2012}, but they become sensitive to systematic errors in the driving Hamiltonian~\cite{Shi2016PRA}.

On the other hand, the technique of ``shortcut to adiabaticity" (STA)~\cite{Berry2009JPA,Chen2010PRL,Campo2013PRL,Giannelli2014RPA} represents an alternative  approach to overcome the longrun time associated with adiabatic evolutions, which is realized by including an auxiliary term to the target Hamiltonian to ``simulate" adiabaticity. The working principle of STA has been demonstrated experimentally in different platforms~\cite{Du2016NatC,Bason2012NatP,Zhang2013PRL,Shuoming2016NatC,Zhou2016np,Zhang2017PRA,Vepsalainenarxiv2017,Huarxiv2018}, including fast quantum state transfer in cold atoms~\cite{Du2016NatC}, acceleration of Bose-Einstein condensate in an optical lattice~\cite{Bason2012NatP}, rapid control of electron spins in nitrogen-vacancy centers~\cite{Zhang2013PRL}, and displacement of trapped ions with minimal excitation~\cite{Shuoming2016NatC}. In addition, experiments have also confirmed the robustness of STA against dissipation and errors~\cite{Du2016NatC,Zhou2016np,Zhang2017PRA}.

\begin{figure}[t]
	\centering
	\includegraphics[width=3.1in,clip=True]{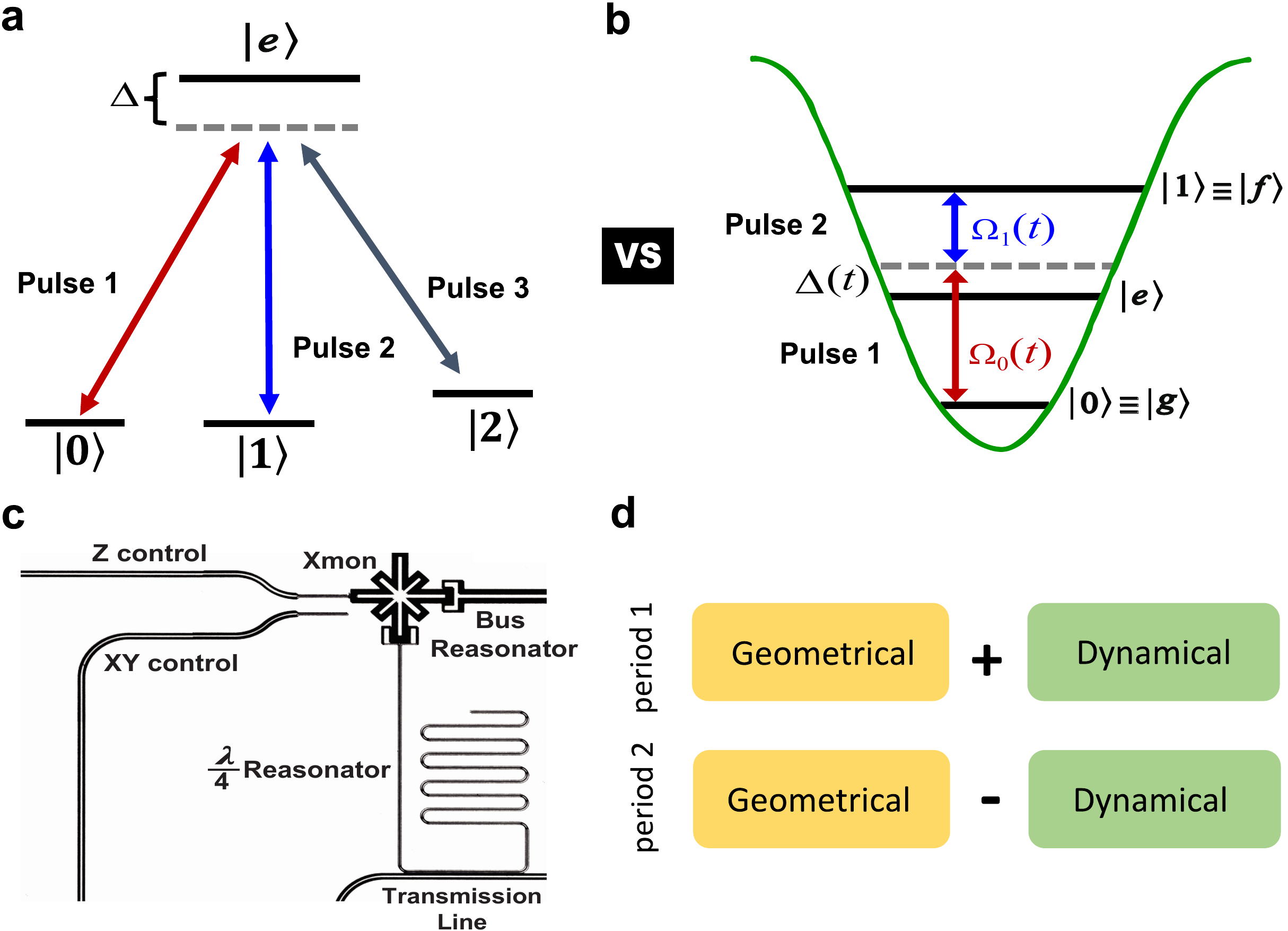}\caption{\label{fig1}\footnotesize{\textbf{Scheme of holomonic quantum computation and qubit structure}.
	\textbf{a.} Coupling scheme of four-level system for non-adiabatic HQC in Ref.~\cite{Duan2001Science,Ditte2007PRA,Gerard2017PRL}. Three pulses couple the ground states $|0\rangle$,$|1\rangle$ and $|2\rangle$ to the excited state $|e\rangle$, and $\Delta$ is the detuning. \textbf{b.} STAHQC scheme using three levels of an Xmon qutrit with a single-photon detuning $\Delta$ as proposed in this work. Two pulses with Rabi frequencies of $\Omega_0(t)$ and $\Omega_1(t)$ are used. \textbf{c.} The Xmon qutrit used in our experiment. Microwave control pulses are imported to the qutrit from the XY control line. The qutrit is dipersively coupled to a $\lambda/4$ resonator for readout, so its state can be inferred from the transmission line output signal $S_{21}$. More details about the sample can be found in Ref.~\cite{CSong2017a}.} \textbf{d.} Elimination of accumulated dynamical phase in our STAHQC gates with two steps using a spin echo pulse.}
\end{figure}

Recently, STA-based techniques have been proposed for realizing robust geometric quantum gates~\cite{Zhang2015Srep,Liang2016PRA,Song2016NJP,Liu2017PRA}. However, existing STA proposals involving non-Abelian geometric phases require applying complicated pulses sequences to simultaneously control four energy levels (Fig.~\ref{fig1}a), making them technologically challenging for an experimental realization and exposing to potentially more sources of control errors. Without a proper  experimental demonstration, the stability of STAHQC gates against NHQC gates can hardly be justified. 

Here we present a theoretical scheme to reduce the complexity in achieving STA holonomic quantum computation (STAHQC), involving the control of only three energy levels instead of four (Fig.~\ref{fig1}b); in this way, all non-Abelian geometric single-qubit gates can be realized in a single non-adiabatic cyclic evolution. Furthermore, the control pulse can be designed beyond the constraints imposed in non-adiabatic HQC (NHQC). Consequently, we can achieve not only a better noise robustness against control errors, but also the capability of pulse optimization. 

For the purpose of demonstration, we report an experimental realization of our proposal using an Xmon superconducting qutrit, which has a ladder $\Xi$ energy structure (Fig.~\ref{fig1}b,c). In the experiment, we constructed non-commutative holonomic gates by varying three independent control parameters to generate the SU(2) transformation group elements, following our STAHQC proposal. The experimental results are in good agreement with our numerical simulations, with both control and environmental noise being taken into account. As a result, both NHQC and STAHQC can now be compared within the same experimental platform.

Before optimization, the performance of  NHQC and STAHQC are on par with each other; this is consistent with the results of a recent experimental demonstration of non-adiabatic HQC using superconducting qubits \cite{ComNat,Abdumalikov2013}. However, for many gates, the approach in Ref.~\cite{Abdumalikov2013} requires at least two cycles to implement, which takes a longer time, making the system more susceptible to environmental noise and control error. In addition, the noise robustness of STAHQC can be further enhanced by pulse optimization as shown in Fig.~\ref{fig2}b-c (see Supplementary Material~\cite{SM} for details). Overall, the advantage of STAHQC over non-adiabatic HQC is expected to be more significant as environmental noise and control error become more prominent. 

Finally, we leave our discussion in analyzing the effect of environmental noise and imperfections in the Supplementary Material~\cite{SM}, where we also present a complete scheme for extending our approach to construct two-qubit STAHQC gates with superconducting qubits, making it possible to implement a universal set of STAHQC gates.

\emph{Setting the stage.}\textbf{--} Let us start with a three-level system, where the ground state $|g\rangle$ and the {\it second} excited state $|f\rangle$ are chosen as logic basis of a qubit, $|0\rangle \equiv |g\rangle$ and $|1\rangle \equiv |f\rangle $, and the first excited state $|e\rangle$ as an auxiliary state. The system is driven by a pair of microwave pulses whose frequencies are detuned from $\omega_{ge}$ or $\omega_{ef}$ by $\Delta(t)$, and have time-dependent amplitudes $\Omega_0(t)$ and $\Omega_1(t)$, and phases $\phi_0 (t)$ and $\phi_1 (t)$ (see Fig. \ref{fig1}b). When the two-photon resonant condition is satisfied~\cite{Chen2012PRA,Giannelli2014RPA,Kumar2016nc,Xu2016nc}, under the rotating-wave approximation, the system Hamiltonian can be written as (with $\hbar \equiv1$): $H_0(t) = \sum_{i=0}^{1}\frac{1}{2}(\Omega_i(t) e^{i\phi_i(t) }| i\rangle\langle e|+ {h.c.} )+\Delta(t)| e\rangle \langle e|$. Let us define a bright state, $| b\rangle \equiv \sin(\frac{\theta}{2})e^{i\phi }| 0\rangle + \cos(\frac{\theta}{2})|1\rangle$, where $\phi \equiv \phi_0(t)-\phi_1(t)$ and $\tan(\theta/2) \equiv \Omega_0(t)/\Omega_1(t)$. We shall keep $\theta$ and $\phi$, hence $\left| b \right\rangle$ to be time independent. The above Hamiltonian can then be expressed as:
\begin{equation}
H_0(t) =\frac{1}{2}(\Omega(t)e^{i\phi—_1(t)}| b\rangle \langle e| +h.c.)+\Delta(t)| e\rangle \langle e|\,
\label{eq1}
\end{equation}
where $\Omega(t) \equiv \sqrt{\Omega_0(t)^2+\Omega_1(t)^2}$ is the Rabi frequency of $H_0(t)$. The instantaneous eigenstates of $H_0(t)$ are $\left| {{E_0}} \right\rangle \equiv |d\rangle =\cos(\frac{\theta}{2})e^{i\phi}| 0\rangle-\sin(\frac{\theta}{2})|1\rangle$, $|E_{+}(t)\rangle = \sin\varphi(t)|b\rangle + \cos\varphi(t)e^{-i\phi—_1(t)}|e\rangle$, and $|E_{-}(t)\rangle = \cos\varphi(t)| b\rangle - \sin\varphi(t)e^{-i\phi—_1(t)}|e\rangle$, where $\varphi(t)$ is defined by $\tan(2\varphi(t)) = \Omega(t)/\Delta(t)$. Note that while the two microwave pulses used for control can be fully specified by their amplitudes ($\Omega_0(t),\Omega_1(t)$) and phases ($\phi_0(t),\phi_1(t)$), the equivalent set of control parameters, namely ($\theta,\phi,\phi_1(t),\varphi(t)$), would be more convenient for our discussion below.

\emph{STA-based holonomic gates.}\textbf{--}The essential idea of STA~\cite{Berry2009JPA} is to include an auxiliary term $H_a(t)$ to the Hamiltonian, $H_{\rm STA}(t)=H_0(t)+H_a(t)$, such that the temporal dynamics of $H_{\rm STA}(t)$ is equivalent to adiabatic evolutions of $H_0(t)$. Specifically, for each eigenstate $\left| {{E_k}\left( 0 \right)} \right\rangle $ of $H_0(0)$, one has ${\mathcal T}{e^{ - i\int_0^T {{H_{{\text{STA}}}}\left( t \right)} dt}}\left| {{E_k}\left( 0 \right)} \right\rangle  = {e^{ - i\int_0^T {{E_k}\left( t \right)dt}  - \int_0^T {\left\langle {{E_k}\left( t \right)} \right.\left| {{{\dot E}_k}\left( t \right)} \right\rangle dt} }}\left| {{E_k}\left( T \right)} \right\rangle $, where for a cyclic evolution, $\left| {{E_k}\left( T \right)} \right\rangle  = \left| {{E_k}\left( 0 \right)} \right\rangle$. Here ${\int_0^T {{E_k}\left( t \right)dt} }$ is the dynamic phase, and ${i\int_0^T {\langle {{E_k}\left( t \right)} | {{{\dot E}_k}\left( t \right)} \rangle dt} }$ is the geometric phase. For our case, the following auxiliary Hamiltonian, $H_a(t)= i \, \dot{\varphi}(t) \, e^{i\phi
_1(t)} \, | b\rangle\langle e|+ {h.c.}$, first obtained in Ref.~\cite{Chen2010PRL}, is employed to construct our 3-level STAHQC gates.

Now, let us consider a cyclic evolution of $H_0(t)$ from $t=0$ to $t=T$. During this interval, the eigenstates are varied in a cyclic fashion, which requires that $\varphi \left( 0 \right) = \varphi \left( T \right)=0$. Additionally, we impose another constraint at the middle, namely $\varphi \left( T/2 \right)=\pi/2$, but $\varphi(t)$ can be varied arbitrarily at other times. In this way, the eigenstate $|E_{-}(t)\rangle$ evolves from $|E_{-}(0)\rangle=|b\rangle$ to $|E_{-}(T/2)\rangle=|e\rangle$, and back to $|E_{-}(T)\rangle=|b\rangle$. On the other hand, the phase $\phi_1(t)$ is varied in the following way: $\phi_1(t)=\gamma_{1}$ for $0\le t \le T/2$, and $\phi_1(t)=\gamma_{2}$ for $T/2< t \le T$, where $\gamma_{1}$ and $\gamma_{2}$ are different constants.

As a result, the geometric phase resulted from such a cyclic evolution, $\gamma  \equiv i\int_0^T {\left\langle {{E_-}\left(t\right)} \right | {{{\dot E}_-}\left(t\right)} \rangle dt}$, is given by $\gamma=\gamma_{1}-\gamma_{2}$. Note that a dynamic phase also accumulates during the evolution, but it can be eliminated with a spin-echo pulse, i.e, a $\pi$-phase shift of the microwave applied halfway ($t=T/2$) of the control sequence (Fig.~\ref{fig1}d; see also the experimental section for details). Furthemore, the dark state $|d\rangle$ is always decoupled from the system, as $H_0\left( t \right)\left| d \right\rangle  = 0$.

Consequently, in the subspace spanned by the two states of $\left| {{E_0}} \right\rangle = \left| d \right\rangle =\cos(\frac{\theta}{2})e^{i\phi}| 0\rangle-\sin(\frac{\theta}{2})|1\rangle$, and $\left| {{E_ - }\left( 0 \right)} \right\rangle  = \left| b \right\rangle = \sin (\frac{\theta }{2}){e^{i\phi }}|0\rangle  + \cos (\frac{\theta }{2})|1\rangle$, the holonomy matrix associated with the above cyclic evolution is given by ${U} = \left| d \right\rangle \left\langle d \right| + {e^{-i\gamma }}\left| b \right\rangle \left\langle b \right|$, which is non-diagonal in the computational basis $\{|0\rangle,|1\rangle\}$,
\begin{equation}
{U}(\theta,\phi,\gamma) = {e^{i\frac{\gamma }{2}}}\left( {\begin{array}{*{20}{c}}
  {{c_{\gamma /2}} - i{s_{\gamma /2}}{c_\theta }}&{ - i{s_{\gamma /2}}{s_\theta }{e^{i\phi }}} \\
  { - i{s_{\gamma /2}}{s_\theta }{e^{ - i\phi }}}&{{c_{\gamma /2}} + i{s_{\gamma /2}}{c_\theta }}
\end{array}} \right) \ ,
\end{equation}
where ${c_x} \equiv \cos x$ and ${s_x} \equiv \sin x$. Alternatively, with $\textbf{n}=(\sin\theta\cos\phi,\sin\theta\sin\phi,\cos\theta)$, we can also write ${U}(\theta,\phi,\gamma) = {e^{i\frac{\gamma }{2}}}{e^{ - i\frac{\gamma }{2}{\mathbf{n}} \cdot \sigma }}$, which describes a rotation around the $\textbf{n}$ axis by a $\gamma$ angle, up to a global phase of $e^{-i\frac{\gamma}{2}}$. Since $\textbf{n}$ and $\gamma$ can be set to any desired values, $U$ can be utilized to construct arbitrary geometric single-qubit gates.

\begin{figure}[tb]
\centering
\includegraphics[width=3.3in,clip=True]{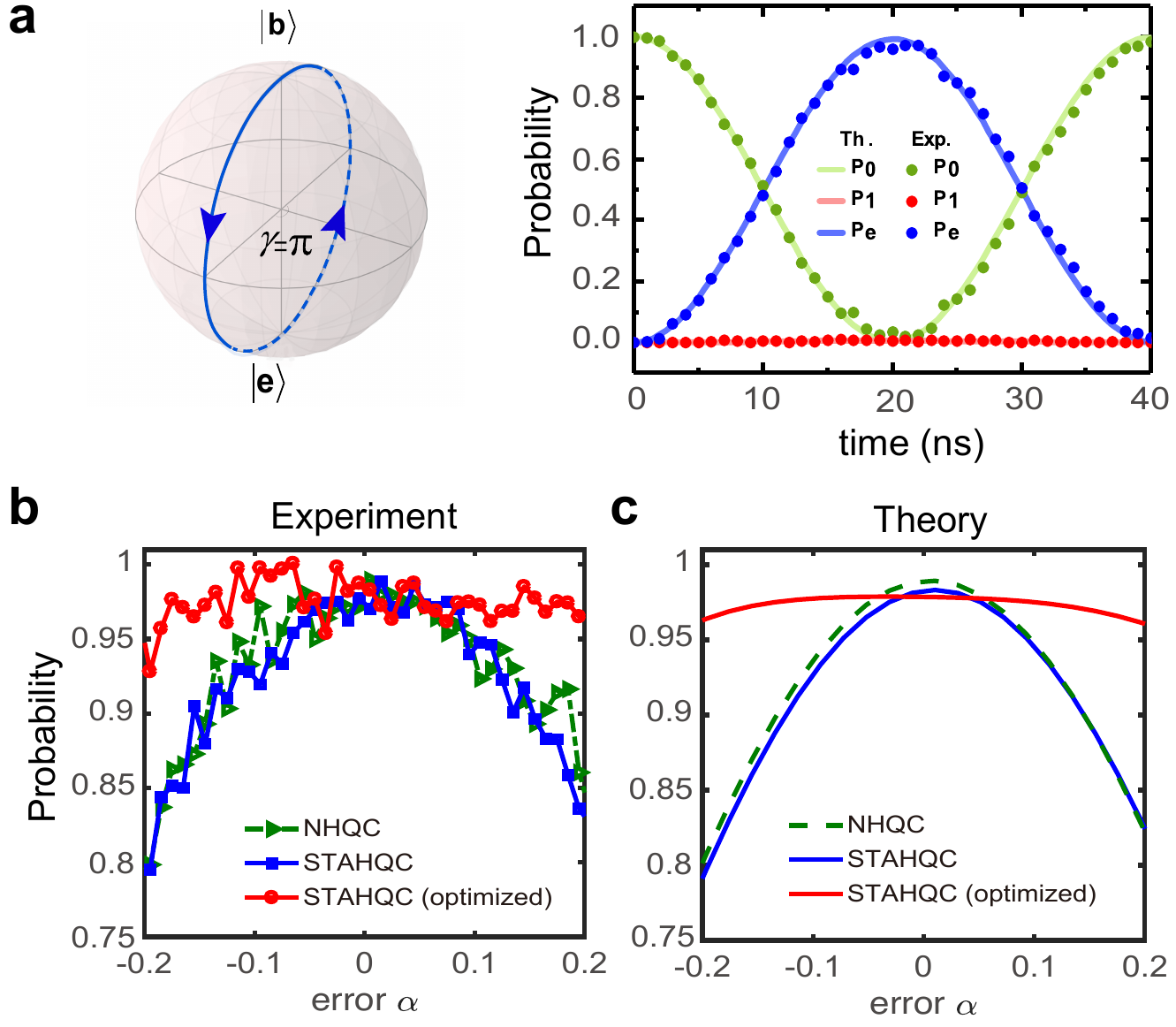}
\caption{\label{fig2}\footnotesize{\textbf{	Evolution of bright state and robustness of holonomic gates.} \textbf{a.} A holonomic $Z$ gate realized by setting $\theta=\pi$, $\phi=0$, and $\gamma=\pi$. In this case, the initial state is the bright state: $|0\rangle=|b\rangle$. It acquires a geometric phase of $\gamma=\pi$ during the gate operation. Experimental results (dot) fit well with numerical simulations (solid lines). \textbf{b} (experiment) and \textbf{c} (theory): performance of an $X$ gate with control errors for various HQC schemes. Theoretical results are obtained using master-equation numerical simulation. $\alpha$ represents magnitude of the control error.}}
\end{figure}

\emph{Experimental results and analysis}.\textbf{--}The three energy levels of our Xmon qutrit are characterized by, $\omega_{ge}/2\pi=5.665$ GHz, and $\omega_{ef}/2\pi =5.417$ GHz. The relaxation and dephasing times of the first and second excited states are $T_1^e=29$ $\mu$s, $T_1^f=9$ $\mu$s, $T_2^{ge}=5.9$ $\mu$s, and $T_2^{ef}=5.8$ $\mu$s, respectively. Level spacing of the qutrit can be fine tuned by a bias current on the Z control line. The control microwave pulses are applied to the qutrit through the XY control line. The qutrit is capacitively coupled to a $\lambda/4 $ resonator ($\omega_{r}/2\pi=6.509$ GHz) with a coupling strength of $ g_{r}/2\pi = 41.3$ MHz, which is in turn coupled to a transmission line. In the dispersive readout scheme~\cite{Jeffrey2014PRL}, the state of the qutrit can be deduced by measuring the transmission coefficient $S_{21}$ of the transmission line. More details about the sample can be found in Ref.~\cite{CSong2017a}.

\begin{figure}[t]
	\centering
\includegraphics[width=3.2in,clip=True]{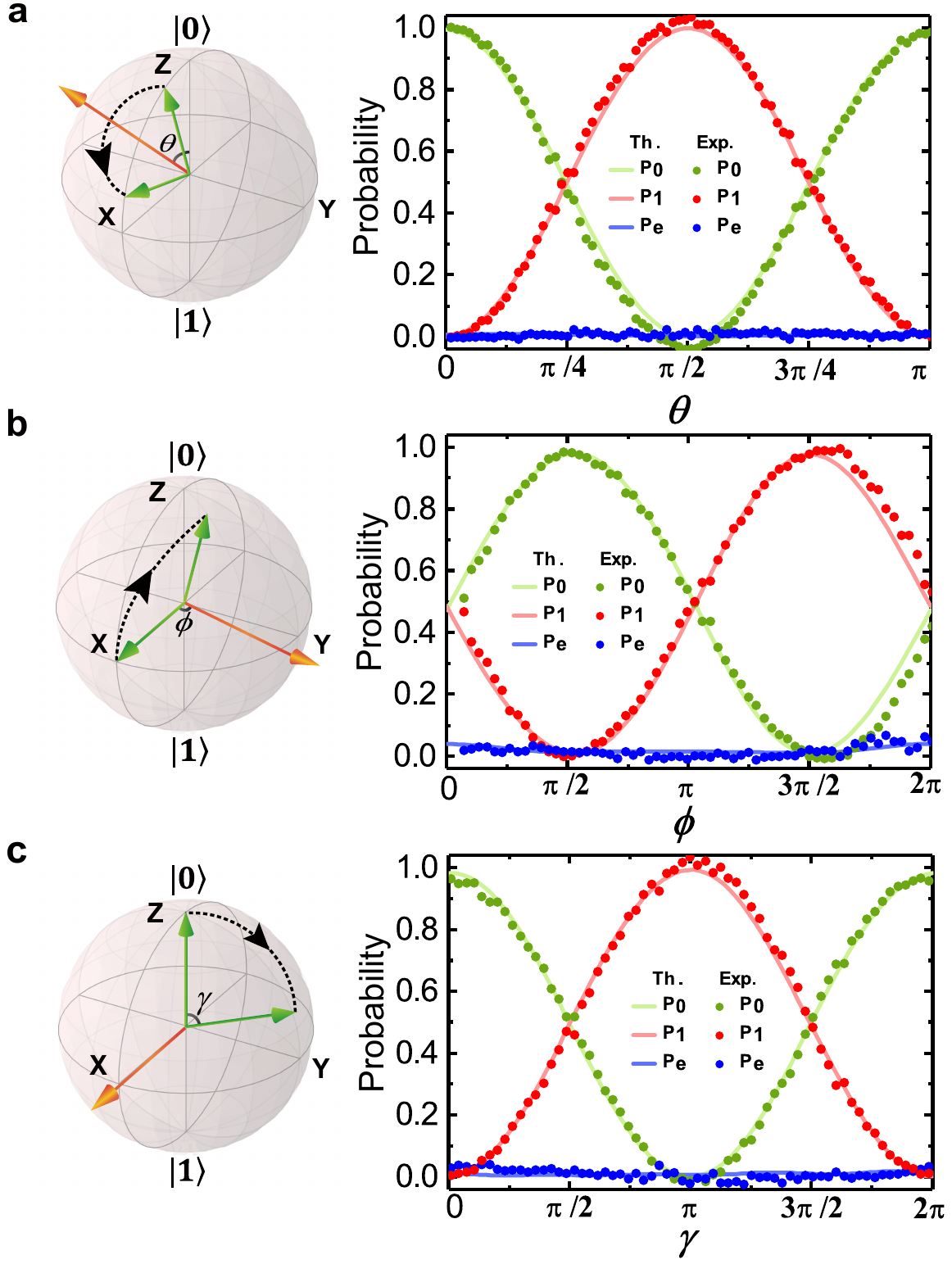}\caption{\label{fig3}\footnotesize{\textbf{Tunable parameters for STAHQC gates.} \textbf{a.} Gates with variable $\theta$ for an initial state of $|0\rangle$. \textbf{b.} Gates with variable $\phi$ for an initial state of $(|0\rangle +|1\rangle)/\sqrt{2}$. \textbf{c.} Gates with variable $\gamma$ for an initial state of $|0\rangle$. Dots and lines are experimental data and numerical simulation, respectively. }}
\end{figure}
\begin{figure*}[t]
	\centering
	\includegraphics[width=7.0in,clip=True]{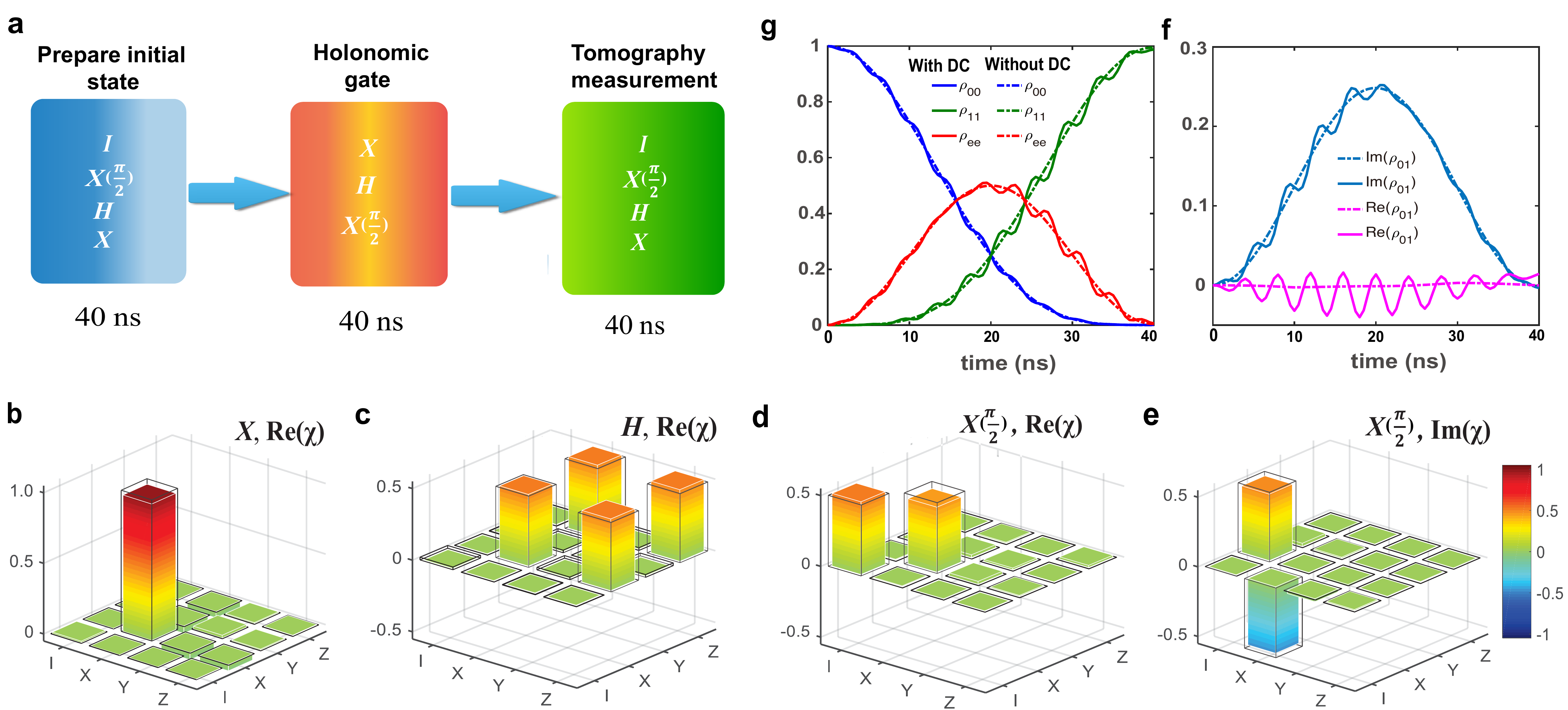}\caption{\label{fig4}\footnotesize{\textbf{Process tomography of holonomic gates}. \textbf{a.} Sequence for process tomography. \textbf{b-e.} Experimental results (colored solid bars) and numerical simulations (black frames) of the tomography matrix $\chi$ for $X$ (\textbf{b}, $(\theta, \phi,\gamma)=(\pi/2,0,\pi)$), $H$ (\textbf{c}, $(\theta, \phi,\gamma)=(\pi/4,0,\pi)$), and $X(\frac{\pi}{2})$ (\textbf{d} and \textbf{e}, $(\theta, \phi,\gamma)=(\pi/2,0,\pi/2)$) gates, respectively. \textbf{g} and \textbf{f}: diagonal and off-diagonal elements of the density matrix for an STAHQC $X$ gate, with (solid lines) and without (dashed lines) of dynamical contribution (DC) of higher energy levels of Xmon. The qubit is initialized into $|0\rangle$. }}
\end{figure*}

We perform a set of gate operations with the following Rabi frequency and detuning: (i) for $ 0 \le t  \le T/2$, $\Omega(t)= \Omega_a \sin(\frac{2\pi t}{T})$ and $\Delta(t)=\Omega_a \cos(\frac{2\pi t}{T})$; (ii) for $T/2 < t  \le T $, $\Omega(t)= -\Omega_a \sin(\frac{2\pi t}{T})$ and $\Delta(t)=-\Omega_a \cos(\frac{2\pi t}{T})$ with $\Omega_a=2\pi\times2$ MHz. With such a choice, the dynamic phases accumulated during $ 0 \le t  \le T/2$ and $ T/2 \le t  \le T$ cancel each other (Fig.~\ref{fig1}d). As a consequence, we are able to realize arbitrary geometrical single-qubit gate by varying the control parameters ($\theta$,$\phi$,$\gamma$).

In the first part, we verify the behavior of bright state. The qubit is initialized to the ground state $|0\rangle$. For the realization of a $Z$ gate, where we set $\theta=\pi,\phi=0,\gamma=\pi$, the ground state is the bright state, i.e., $\left| b \right\rangle  = \left| 0 \right\rangle$. It thus follows the evolution of $\left| {{E_ - }\left( t \right)} \right\rangle$ without transition to other states (see Fig. \ref{fig2}{a}).

To demonstrate that an arbitrary SU(2) transformation can be achieved using our method, we experimentally verify that all three parameters, $\theta,\phi$, and $\gamma$, can be varied continuously and independently. We first apply the following gate, $U(\theta,\phi=0,\gamma=\pi)$, to an initial state of $|0\rangle$, and investigate the final state as a function of $\theta$, i.e., $U(\theta,0,\pi)|0\rangle=\cos(\theta/2) |0\rangle+\sin(\theta/2)|1\rangle$. This gate operation corresponds to a rotation along the axis $\textbf{n}=(\sin\theta,0,\cos\theta)$ by an angle of $\gamma=\pi$ (see Fig.~\ref{fig3}a). Next, we apply the gate $U(\pi/2,\phi,\pi/2)$ to an initial state of $(|0\rangle+|1\rangle)/\sqrt{2}$, which corresponds to a rotation along the axis $\textbf{n}=(\cos\phi,\sin\phi,0)$ by an angle of $\pi/2$, as shown in Fig.~\ref{fig3}b. Finally, we demonstrate that the geometric phase $\gamma$ is also continuously tunable by applying the gate $U(\pi/2,0,\gamma)$ to an initial state of $|0\rangle$. This gate operation is essentially a rotation along the $x$-axis by an angle of $\gamma$, as plotted in Fig.~\ref{fig3}c.

Overall, the fidelity of the quantum gate is characterized by quantum process tomography. Since our STA scheme can generate arbitrary single-qubit gates, we may use them throughout the complete process tomography, including (i) initial-state preparation, (ii) quantum-gate implementation, as well as (iii) final-state rotation for state tomography. For example, Fig.~\ref{fig4}a shows the sequence of quantum process tomography using the following set of gates, $I$, $X(\pi/2)$, $H$, and $X$, to generate four initial states, $|0\rangle$, $(|0\rangle + i|1\rangle)/\sqrt{2}$, $(|0\rangle+|1\rangle)/\sqrt{2}$, and $|1\rangle$, and investigate fidelity of the gates $X$, $H$, and $X(\pi/2)$. The experimental results are shown in Fig.~\ref{fig4}b-e. The process fidelities for $X$, $H$, and $X(\pi/2)$ are $F_X=96.6\pm 0.8 \%$, $F_{H}=97.6\pm 1.0\%$, and $F_{X(\frac{\pi}{2})}= 96.4\pm 1.0 \%$, respectively. Numerical simulations using a master equation method, taking into account of dissipation, give fidelities of $F_X=98.4\%$, $F_{H}=98.4\%$, and $F_{X(\frac{\pi}{2})}= 97.8\%$, which are in good agreement with the experimental results. The major sources of error include dynamical contribution from higher energy levels, decoherence, and control pulse errors (see Ref.~\cite{SM}), shown in Fig.~\ref{fig4}g,f. In principle, the dynamical contribution from higher energy levels can be much suppressed by using highly nonlinear systems such as flux qubits, whereas the decoherence issue can be improved by using qubits with longer $T_1$ and $T_2$ times. As for the effect of control pulse error, in the Supplementary Material~\cite{SM} we specifically show, both numerically and experimentally, that our STAHQC gates exhibit a better robustness than NHQC.

In summary, we have proposed and experimentally demonstrated single-looped holonomic gates based on the technique of shortcut to adiabaticity, which is robust against control errors and environmental noise~\cite{Chen2012PRA,Du2016NatC,Zhou2016np}. Our STAHQC approach is compatible with other optimization methods~\cite{DaemsPRL2013,Hong2017arX,Egger2018arXiv} for further enhancement of gate fidelity. Moreover, it can be extended to construct two-qubit holonomic gates to realize a universal STAHQC gate set, as we explicitly present how to construct a two-qubit $\sqrt{SWAP}$ gate in the Supplementary Material~\cite{SM}. This method should also be of interest to other platforms such as nitrogen-vacancy centers, trapped ions, quantum dots, and nuclear magnetic resonance, etc.

{\it Note added--} After this work was completed, the theoretical idea related to STAHQC had been largely expanded~ \cite{liu2018arXiv} by some of the current authors. Furthermore, a recent experiment has been reported demonstrating an improvement on NHQC~\cite{xu2018arXiv}. However, the sensitivity to systematic noise has not been improved.  

This work was supported by Natural Science Foundation of Guangdong Province (2017B030308003), the Guangdong Innovative and Entrepreneurial Research Team Program (No.2016ZT06D348), and the Science Technology and Innovation Commission of Shenzhen Municipality (ZDSYS20170303165926217, JCYJ20170412152620376). We particularly thank Prof. Haohua Wang at Zhejiang University, where all the experimental data were taken, for providing access to the experimental facilities, as well as his valuable discussions and comments on the manuscript.

\end{document}